\documentclass[journal]{IEEEtran}

\usepackage{cite}      

\usepackage{graphicx}  

\usepackage{amsmath,amssymb}   
\begin{document}
%
\title{Control of ferroelectric phase by chemical pressure in (Gd,Tb)MnO$_{3}$ crystals}
%
%

\author{K. Noda, S. Nakamura, and H. Kuwahara\thanks{K. Noda, S. Nakamura, and H. Kuwahara are with the Department of Physics, Sophia University, 7-1 Kioi-cho, Chiyoda-ku, Tokyo, 102-8554, Japan. E-mail: n-kohei@sophia.ac.jp.}} 

%
%
%
\markboth{Journal of \LaTeX\ Class Files,~Vol.~1, No.~11,~November~2002}{Shell \MakeLowercase{\textit{et al.}}: Bare Demo of IEEEtran.cls for Journals}
%



\maketitle

\begin{abstract}
We have investigated the dielectric properties of Gd$_{1-x}$Tb$_{x}$MnO$_{3}$ crystals in magnetic fields to clarify the crossover between two distinct ferroelectric phases, GdMnO$_{3}$-type (first-order displacive) and TbMnO$_{3}$-type (second-order order-disorder). 
We have found that the compositional phase boundary between the phases exists around 0.3$<$$x$$<$0.4. 
For the $x$=0.1 sample exposed to magnetic fields above 3T, we have discovered successive ferroelectric phase transitions that occur at two different temperatures.
We have determined the electric and magnetic phase diagram for $x$$=$0, 0.1, and 0.3 in the plane of temperature and field in order to demonstrate the change in the ferroelectric phases. 
\end{abstract}

\begin{keywords}
Electromagnetic coupling, Ferroelectric materials, Antiferromagnetic materials.
\end{keywords}

%
\IEEEpeerreviewmaketitle

\section{Introduction}
%
%
%
%
\PARstart{M}{ultiferroic} materials, which are simultaneously (anti)ferromagnetic and (anti)ferroelectric, are now the subject of renewed interest \cite{kimura,Kuwahara,Noda}.
However, in spite of intensive recent research, such substances are quite rare. 
A newly discovered compound, TbMnO$_{3}$, is a multiferroic material. 
In zero magnetic field, it shows a spontaneous ferroelectric polarization along the $c$ axis ($P_{c}$$^{{\rm F}}$) below the incommensurate-commensurate (IC-C) or lock-in antiferromagnetic (AF) transition, temperature ($T_{\rm{lock}}$) \cite{kimura}. 
In magnetic fields along the $b$ axis, its direction of ferroelectric polarization ($P^{{\rm F}}$) changes from along the $c$ to the $a$ axis, an event called a ``magnetic-field-induced electric polarization flop''. 
The compound TbMnO$_{3}$ has several magnetic transitions: the IC AF transition ($T_{{\rm IC}}$$\sim$41K), the IC-C (lock-in) AF transition ($T_{{\rm lock}}$$\sim$27K), and the ordering transition of the Tb 4$f$ moments ($T^{\rm Tb}$$\sim$7K). 
The ferroelectric transition of TbMnO$_{3}$ corresponds to an IC-C AF transition that is due to the magnetic ordering of Mn 3$d$ spins. 

In the case of another such compound, GdMnO$_{3}$ \cite{Kuwahara}, we have found the ferroelectric phase below 13K. 
At this temperature, the Gd 4$f$ moments antiferromagnetically couple with the Mn 3$d$ spins and the weak ferromagnetism of the Mn 3$d$ spins, caused by the Dzyaloshinskii-Moriya (DM) interaction, is suppressed. 
Transitions displayed by GdMnO$_{3}$ include the IC AF transition ($T_{{\rm IC}}$$\sim$42K), the $A$-type AF transition ($T_{{\rm N}}^{\rm Mn}$$\sim$20K), and the ordering transition of the Gd 4$f$ moments or an AF coupling between the Gd 4$f$ moments and Mn 3$d$ spins ($T_{{\rm N}}^{\rm Gd}$$\sim$13K). 
In GdMnO$_{3}$, the $A$-type AF transition and the transition of the Gd 4$f$ moments possess a thermal hysteresis, and the temperatures listed above refer to a warming scan. 
The $P^{{\rm F}}$ of GdMnO$_{3}$, relative to the order of the Gd 4$f$ moments, exists along the $a$ axis in zero field \cite{Kuwahara}. 
In our previous work, we demonstrated that the mechanism of the ferroelectric transition in GdMnO$_{3}$ is different from that in TbMnO$_{3}$ in zero field, although Gd and Tb ions are adjoining members of the lanthanum series \cite{Noda}. 
The former should be classified as a first-order displacive-type transition, and the latter as a transition of the second-order order-disorder-type. 
The difference in ferroelectric transition mechanisms between GdMnO$_{3}$ and TbMnO$_{3}$ are thought to originate in their degree of orthorhombic distortion and in the magnitude of their rare earth moments. 
Therefore, to clarify the crossover between the two distinct magnetic and ferroelectric phases, we measured the ferroelectric and magnetic properties of $R$MnO$_{3}$ ($R$$=$Gd$_{1-x}$Tb$_{x}$, 0$\le$$x$$\le$1) compounds while exerting a fine degree of control of the degree of orthorhombic distortion, also known as the chemical pressure, and the magnitude of the rare earth moments, by changing $R$-site chemical composition.


 




\section{Experiment}
All single crystals used in this work were prepared by the floating zone method. 
We carried out an x-ray diffraction experiment to confirm that all crystals showed an orthorhombic $Pbnm$ structure without any impurity phase. 
All specimens used in the measurements were cut along the crystallographic principal axes into a rectangular shape.
The temperature dependence of the dielectric constant ($\varepsilon$) and $P^{{\rm F}}$ were measured in a temperature-controllable cryostat with a superconducting magnet that provided a field up to 8T\@. 
The $P^{{\rm F}}$ was estimated by a pyroelectric-current measurement taken in a warming scan at a rate of 4K/min, after the sample had been cooled under a poling field of 300$\sim$500kV/m. 

\section{Results and Discussion}

\begin{figure}
\centering
\includegraphics[scale=0.31]{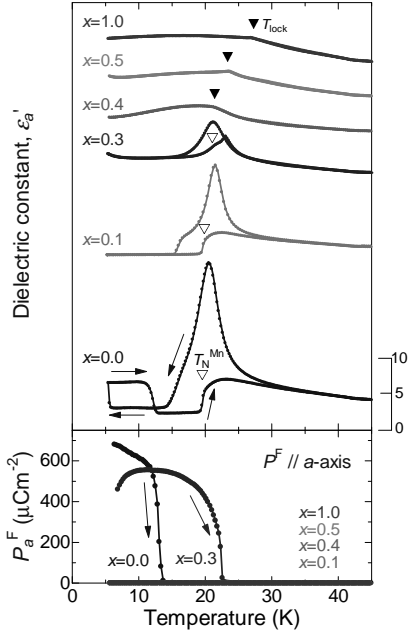}
\vspace{-3mm}
\caption{Temperature dependence of the dielectric constant along the $a$ axis ($\varepsilon$$_{a}$ (upper panel)) and $P^{{\rm F}}$ along the same axis (lower) for Gd$_{1-x}$Tb$_{x}$MnO$_{3}$ (0.0$\le$$x$$\le$1.0) crystals in zero field. The AF transition temperature $T_{{\rm N}}$$^{{\rm Mn}}$ and $T_{{\rm lock}}$ are shown by open and solid triangles, respectively. The arrows indicate the direction of temperature scan. For clarity, the plot for each compound in the upper panel is shifted vertically without any change of scale (refer to the scale shown in the right side of the panel).} 
\label{fig1}
\end{figure}

Figure \ref{fig1} shows the dielectric and ferroelectric properties of several Gd$_{1-x}$Tb$_{x}$MnO$_{3}$ (0.0$\le$$x$$\le$1.0) crystals. 
The amount of orthorhombic distortion was systematically increased by varying $x$ between 0 and 1. 
As a result, the dielectric behavior systematically changed from GdMnO$_{3}$-like to TbMnO$_{3}$-like. 
It is clear that, the dielectric constant along the $a$ axis ($\varepsilon$$_{a}$) changed drastically from GdMnO$_{3}$-type to TbMnO$_{3}$-type around 0.3$<$$x$$<$0.4. 
A sharp change in $\varepsilon$$_{a}$, associated with the AF transition temperature $T_{{\rm N}}$$^{{\rm Mn}}$, that exhibited thermal hysteresis ($\triangledown$ symbol in Fig.\ref{fig1}) gradually disappeared and changed to a small inflection associated with $T_{{\rm lock}}$ as $x$ increased. 
At the same composition (0.3$<$$x$$<$0.4) the direction of $P^{{\rm F}}$ changes from along the $a$ axis to the $c$ axis. 
With increasing $x$, the anomaly indicated by $\triangledown$ symbol below $x$$=$0.3 and that by $\blacktriangledown$ symbol above $x$$=$0.4 shifted toward higher temperature. 

Except for the two end members, the magnetic structures of these materials below the temperature of the anomaly have not yet been identified. 
However, the systematic change in $\varepsilon$$_{a}$ suggest that the anomalies indicated by the open and solid triangles correspond to $T_{{\rm N}}$$^{{\rm Mn}}$ and $T_{{\rm lock}}$, respectively. 
Concerning our initial motivation of clarifying the change of dielectric properties between $x$$=$0 and $x$$=$1, we can conclude that the compositional phase boundary between $x$$=$0 (GdMnO$_{3}$-type, first-order displacive-type) and $x$$=$1 (TbMnO$_{3}$-type, second-order order-disorder-type) exists around 0.3$<$$x$$<$0.4.

For the $x$$=$0.1 sample in zero field, the ferroelectric polarization along the $a$ axis ($P_{a}$$^{{\rm F}}$, FE1) disappeared (see lower panel), and the thermal hysteresis in $\varepsilon$$_{a}$ due to the ferroelectric transition below 13K, as seen in the $x$$=$0 sample, was also suppressed (upper panel). 
However, when we increased $x$ to 0.3, another $P_{a}$$^{{\rm F}}$ (FE2) was observed at around $T_{{\rm N}}$$^{{\rm Mn}}$. 
The transition character of FE2 seems to be different from that of GdMnO$_{3}$ (FE1). 
In the case of GdMnO$_{3}$, the origin of FE1 is the magnetic ordering of Gd 4$f$ moments and the ferroelectric transition is accompanied by thermal hysteresis, as described in the introduction. 
Therefore, unless a sample is cooled below 6K, at which temperature the magnetic transition of the Gd 4$f$ moments occurs in a cooling process \cite{Hamberger}, FE1 does not appear at all. 
In order to clarify the difference between FE1 and FE2, we have investigated the existence of $P^{{\rm F}}$ (FE2) for the $x$$=$0.3 sample after the sample was cooled down to several different temperature above 6K. 
Consequently, it turns out that the $P^{{\rm F}}$ (FE2) still appears even if the sample is cooled down to 10K, a temperature higher than the magnetic transition temperature of rare-earth ion 4$f$ moments. 
In the case of the Gd ion, the magnetic transition temperature is 6K \cite{Hamberger}, and that of Tb ion is 7K \cite{Kajimoto}.
This result strongly suggests that FE2 has a different origin from FE1 and is associated with the ordering of the Gd 4$f$ moments. 
Therefore, we propose that the origin of FE2 is related to $T_{{\rm N}}$$^{{\rm Mn}}$, which is the temperature of an AF transition of Mn 3$d$ spins. 
We investigated the lattice striction of this compound and there was no first-order like transition at $T_{{\rm N}}$$^{{\rm Mn}}$.
This result suggested that the ferroelectric transition of FE2 in $x$=0.3 is the order-disorder type, like TbMnO$_{3}$. 
With further increase of $x$, FE2 disappeared and $P_{c}$$^{{\rm F}}$ (signaling the appearance of the TbMnO$_{3}$-type ferroelectric phase) appeared for the $x$$=$0.4 sample. (not shown in Fig.\ref{fig1}.)

\begin{figure}
\centering
\includegraphics[scale=0.31]{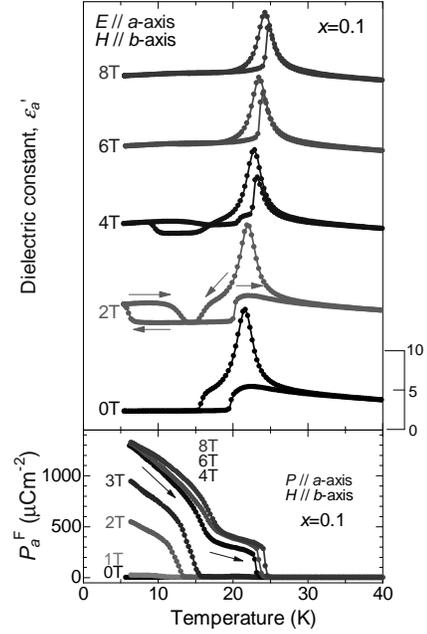}
\vspace{-3mm}
\caption{Ferroelectric transition induced by application of a field for the Gd$_{1-x}$Tb$_{x}$MnO$_{3}$ $x$$=$0.1 sample. Upper (lower) panel shows the temperature dependence of $\varepsilon$$_{a}$ ($P_{a}$$^{{\rm F}}$) in magnetic fields applied parallel to the $b$ axis. The arrows indicate the direction of $T$ scan. For clarity, the plot for each fields in upper panel is shifted vertically without any change of scale (refer to the scale shown in the right side of the panel). }
\label{fig2}
\end{figure}

Next, in order to reveal the crossover between FE1 ($x$$=$0) and FE2 ($x$$=$0.3), we examined the dielectric (upper panel of Fig.\ref{fig2}) and ferroelectric (lower panel) properties of the $x$$=$0.1 sample, in which $P^{{\rm F}}$ can not be detected along any axis $in$ $zero$ $field$, in various magnetic fields applied parallel to the $b$ axis. 
We chose this direction to enhance the ferroelectric phase (FE1) because the $b$ axis is the easy magnetic axis in the case of $x$$=$0 (GdMnO$_{3}$) \cite{Kuwahara,Noda}. 
The behavior of $\varepsilon$$_{a}$ for $x$$=$0.1 changes from GdMnO$_{3}$-like (FE1) to FE2-type behavior at $x$$=$0.3 with increase of field. 
In spite of absence of a ferroelectric phase $in$ $zero$ $field$, when we applied fields up to 2T in strength parallel to the $b$ axis, thermal hysteresis in $\varepsilon$$_{a}$ similar to the case of $x$$=$0 appeared below 13K, and the $P_{a}$$^{{\rm F}}$ also appeared. 
Furthermore, if the field was increased to more than 4T, a peak emerges in $\varepsilon$$_{a}$ around $T_{N}$$^{{\rm Mn}}$ in the warming process. 
As can be clearly seen in the lower panel of Fig.\ref{fig2}, the transition temperature for the appearance of $P_{a}$$^{{\rm F}}$, corresponding to the dielectric peak, systematically increased for fields above 4T. 
We suppose that the ferroelectric transition temperature ($T_{{\rm C}}$) of the $x$$=$0.1 compound $in$ $zero$ $field$ in a cooling scan occurs at lower temperature than we could attain.
In other words, by the application of the fields the $T_{{\rm C}}$ was shifted toward a higher temperature that we could observe. 
Thus, we suggest that the ferroelectric phase in the $x$$=$0.1 material present below 3T has the same origin as FE1 in the $x$$=$0 compound (GdMnO$_{3}$). 

On the other hand, $in$ $magnetic$ $field$ $above$ $4T$, the $P^{{\rm F}}$ that appeared around $T_{N}$$^{{\rm Mn}}$ ($\sim$23K) closely resembles FE2 in the $x$$=$0.3 sample. 
As one can immediately notice from the lower panel of Fig.\ref{fig2}, $P_{a}$$^{{\rm F}}$ $in$ $the$ $field$ $above$ $4T$ shows two successive phase transitions: an abrupt jump around $T_{{\rm N}}$$^{{\rm Mn}}$ ($\sim$23K) corresponding to FE2 and a gradual increase around 15K corresponding to FE1. 
These results suggest that both FE1 and FE2 coexist in the $x$$=$0.1 sample $in$ $the$ $higher$ $fields$. 

We investigated the cooling-hysteresis dependence of $P^{{\rm F}}$ in the $x$$=$0.1 compound $in$ $fields$ for the purpose of exploring the possible existence of FE1 and$/$or FE2 and the correlation between FE1 and FE2 (i.e. the same zero field experiment as discussed previously for the 0.3 sample). 
After cooling down to 18K under 4T, we observed only the FE2-like $P_{a}$$^{{\rm F}}$. 
This result strongly implies that both FE1 and FE2 coexist independently, although transitions of FE1 and FE2 occurred sequentially for $x$$=$0.1 (Gd$_{0.9}$Tb$_{0.1}$MnO$_{3}$) $in$ $a$ $field$.

\begin{figure}
\includegraphics[scale=0.36]{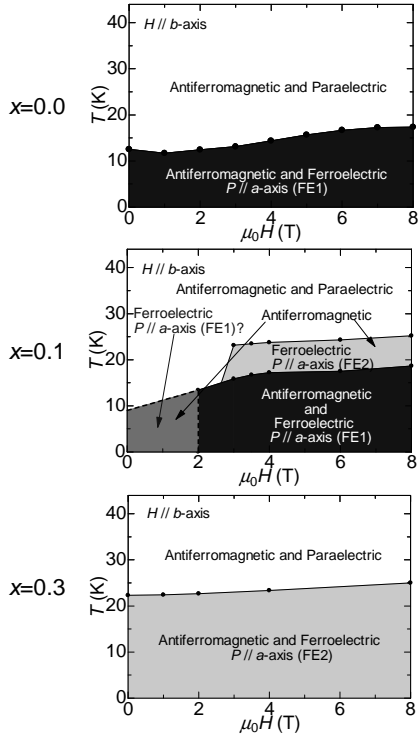}
\vspace{-3mm}
\caption{Electric and magnetic phase diagrams of Gd$_{1-x}$Tb$_{x}$MnO$_{3}$ ($x$=0.0(top), 0.1(middle), and 0.3(bottom)) in the plane of temperature and field. The $x$=0.1 sample shows the presence of both FE1 and FE2 simultaneously at higher magnetic fields (see text). The dotted line between 0$\le$$x$$\le$2T was extrapolated from the value of higher magnetic fields.
 }
\label{fig3}
\end{figure}

To better illuminate this complicated situation, we have summarized our results in the electric and magnetic phase diagrams shown in Fig.\ref{fig3} for Gd$_{1-x}$Tb$_{x}$MnO$_{3}$ ($x$=0.0, 0.1, and 0.3). 
These are derived from the dielectric anomalies observed in fields parallel to the $b$ axis. 
In the case of $x$$=$0 (GdMnO$_{3}$) (top panel), FE1 was observed in fields up to 8T and it was enhanced by application of the field. 
This result indicates that $T_{{\rm C}}$ increases with the field strength. 
On the other hand, $x$$=$0.3 (bottom panel) shows FE2 in fields, which was also enhanced at higher fields. 
This behavior seems to be connected with the stabilizing AF ordering of Mn 3$d$ spins. 
In the $x$$=$0.1 case (middle panel), we did not observe $P^{{\rm F}}$ $in$ $zero$ $field$.
However, we believe that, for this material, the $T_{{\rm C}}$ occurs below 5K, the lowest temperature we can attain. 
Accordingly, we have represented the low temperature region of $x$$=$0.1 below 2T as FE1. 
We have confirmed the existence of FE1 in $x$$=$0.1 in a field of 2T. 
When we increased the fields up to 3T, FE2 appeared at a higher temperature than FE1, and both FE1 and FE2 coexisted in fields up to 8T. 
This phase assignment is reasonable considering the results of a systematic increase of $T_{{\rm C}}$ with the fields for $x$$=$0 (FE1) and 0.3 (FE2).

\section{Conclusions}
We have investigated the systematic variation of dielectric and ferroelectric properties between GdMnO$_{3}$ and TbMnO$_{3}$ in magnetic fields. 
We have clearly demonstrated that the compositional phase boundary between $P_{a}$$^{{\rm F}}$ (GdMnO$_{3}$-type: FE1) and $P_{c}$$^{{\rm F}}$ (TbMnO$_{3}$-type) is located at 0.3$<$$x$$<$0.4. 
Moreover, we have found that another ferroelectric phase (FE2) exists even below the phase boundary, which was contrary to our simple expectation, i.e. the ferroelectric transition would systematically change from GdMnO$_{3}$ to TbMnO$_{3}$. 
Although FE2 originates in the AF ordering of Mn 3$d$ spins seen in TbMnO$_{3}$-type materials, the direction of $P^{{\rm F}}$ is not along the $c$ axis but is along the $a$ axis, as observed in GdMnO$_{3}$-type compounds. 
These results suggest that $P^{{\rm F}}$ with the same origin (the AF ordering of Mn 3$d$ spins) can be made to occur along the $a$ or the $c$ axis simply by changing the orthorhombic distortion and the magnetic moment of the 4$f$ electrons. 
We have found successive phase transitions for $x$$=$0.1, in which both FE1 and FE2 coexisted $in$ $a$ $magnetic$ $field$. 
The wide variety of ferroelectric phases found as a function of field or change in orthorhombic distortion and 4$f$ moments will likely provide significant information on the mechanism of the ferroelectric phase transition in this system.

\end{document}